\documentclass[aps,prd,reprint,showpacs,showkeys,superscriptaddress]{revtex4-1}
\usepackage{amsmath}
\usepackage{amssymb}
\usepackage{bm}
\usepackage[pdftex,colorlinks,hyperindex,plainpages=false,bookmarksopen,bookmarksnumbered,pdfusetitle]{hyperref}
\usepackage{mathrsfs}
\usepackage{times}

\DeclareMathOperator\tr{Tr}
\newcommand\dd{\text{d}}
\newcommand\g{\mathfrak{g}}
\newcommand\h{\mathfrak{h}}
\newcommand\imag{\text{i}}
\newcommand\La{\mathscr{L}}
\newcommand\gr[1]{\text{#1}}
\newcommand\gtr[2]{\mathcal T_{#1}{#2}}

\begin{document}

\title{Topological interactions of Nambu--Goldstone bosons in quantum many-body systems}

\author{Tom\'{a}\v{s} Brauner}
\email{brauner@hep.itp.tuwien.ac.at}
\affiliation{Institute for Theoretical Physics, Vienna University of Technology, Vienna, Austria}
\affiliation{Department of Theoretical Physics, Nuclear Physics Institute of the ASCR, \v Re\v z, Czech Republic}

\author{Sergej Moroz}
\email{morozs@uw.edu}
\affiliation{Department of Physics, University of Colorado, Boulder, CO 80309, USA}
\affiliation{Department of Physics, University of Washington, Seattle, WA 98195, USA}

\begin{abstract}
We classify effective actions for Nambu--Goldstone (NG) bosons assuming absence of anomalies. Special attention is paid to Lagrangians invariant only up to a surface term, shown to be in a one-to-one correspondence with Chern--Simons (CS) theories for unbroken symmetry. Without making specific assumptions on spacetime symmetry, we give explicit expressions for these Lagrangians, generalizing the Berry and Hopf terms in ferromagnets. Globally well-defined matrix expressions are derived for symmetric coset spaces of broken symmetry. The CS Lagrangians exhibit special properties, on both the perturbative and the global topological level. The order-one CS term is responsible for non-invariance of canonical momentum density under internal symmetry, known as the linear momentum problem. The order-three CS term gives rise to a novel type of interaction among NG bosons. All the CS terms are robust against local variations of microscopic physics.
\end{abstract}

\maketitle


\section{Introduction}
\label{sec:intro}

The low-energy physics of many-body systems is dominated by collective modes of their elementary constituents, such as sound waves in solids and fluids, spin waves in (anti)ferromagnets, or Bogoliubov modes in superfluids. As a rule, these can be viewed as Nambu--Goldstone (NG) bosons of spontaneously broken continuous symmetries of the system. The broken symmetries are most conveniently encoded in a local effective field theory (EFT) for the NG modes~\cite{Burgess:1998ku,*Brauner:2010wm}.

Terms of topological origin are ubiquitous in quantum field theories for a vast range of physical systems. In high-energy physics, a Wess--Zumino (WZ) term is responsible for anomalous interactions of pions~\cite{Wess:1971yu}. In condensed-matter physics, topological actions play a decisive role for the quantum Hall effect, the dynamics of spin chains, superconductors, topological insulators and other intriguing phenomena~\cite{fradkin2013field,Altland:book}.

Here and in the companion paper~\cite{EFTpaper}, we give a systematic construction of EFTs for NG bosons in the gradient expansion, based on the strategy outlined in Ref.~\cite{Leutwyler:1993iq}. In the present paper, we focus on quasi-invariant Lagrangians, that is, those invariant up to a surface term. Despite intensive research of NG bosons in quantum many-body systems~\cite{Watanabe:2012hr,*Hidaka:2012ym,*Takahashi:2014vua,Watanabe:2013uya,Nicolis:2012vf,*Nicolis:2013sga,*Nicolis:2013lma,*Griffin:2013dfa,*Geracie:2014iva,*Delacretaz:2014jka,*Kampf:2014rka,*Hinterbichler:2014cwa,Takeuchi:2013mwa,*Hayata:2013vfa,*Brauner:2014aha,*Kobayashi:2014xua,*Kobayashi:2014eqa,*Watanabe:2014zza}, explicit expressions for quasi-invariant Lagrangians have only been known for a few particular cases of interest. One of our main results here is a complete classification, and an explicit derivation, of such terms. The explicit solution for the leading-order Lagrangian~\cite{Watanabe:2013uya,Watanabe:2014fva}, which took two decades since its original formulation~\cite{Leutwyler:1993gf}, follows as a simple special case. 

For internal symmetries characterized by a compact Lie group, quasi-invariant Lagrangians are in a one-to-one correspondence with generators of de Rham cohomology groups of the coset space of broken symmetry~\cite{DHoker:1994ti}. In four-dimensional Lorentz-invariant systems, they invariably signal anomalous microscopic dynamics, and can in principle be constructed using differential-geometric methods~\cite{Hull:1990ms,*DHoker:1995it,*DHoker:1995ek,*deAzcarraga:1997gn,*deAzcarraga:1998bu}. We show that in some many-body systems, presence of quasi-invariant Lagrangians does not require the broken symmetry to be anomalous. \emph{Assuming} absence of anomalies, we construct all quasi-invariant Lagrangians using only elementary field theory, without any assumptions on spacetime geometry. These Lagrangians can be mapped to Chern--Simons (CS) theories for unbroken symmetry. Their topological nature is manifested by robustness against local variations of microscopic physics, and tension between manifest locality and gauge invariance.


\section{Gauge-invariant actions}
\label{sec:actions}

Consider a system with a continuous \emph{internal} symmetry group $G$, spontaneously broken to $H\subset G$. Its low-energy physics can be probed by coupling the conserved currents of $G$ to a set of background gauge fields, $A^i_\mu(x)$. It is captured by an EFT, defined by the action $S_\text{eff}\{\pi,A\}$, where $\pi^a(x)$ is a set of NG fields, one for each broken generator $T_a$~\footnote{We adopt the following notation: $T_{i,j,\dotsc}$ for all generators of $G$, $T_{\alpha,\beta,\dotsc}$ for generators of the unbroken subgroup $H$, and $T_{a,b,\dotsc}$ for the broken generators.}. In the absence of anomalies and upon a suitable choice of the variables $\pi^a$, $S_\text{eff}\{\pi,A\}$ becomes invariant under a simultaneous gauge transformation of the NG and background fields~\cite{Leutwyler:1993iq}. The latter reads $\gtr{\g}{A_\mu}\equiv\g A_\mu\g^{-1}+\imag\g\partial_\mu\g^{-1}$, where $\g\in G$ and $A_\mu\equiv A^i_\mu T_i$. The action of symmetry on the NG fields is defined by treating them as coordinates on the coset space $G/H$~\cite{Coleman:1969sm,*Callan:1969sn}. They are encoded in a matrix $U(\pi)$ in some faithful representation of $G$, and their transformation rule reads
\begin{equation}
U(\pi'(\pi,\g))=\g U(\pi)\h(\pi,\g)^{-1},
\label{pitransfo}
\end{equation}
where $\h\in H$. With the choice $\g=U(\pi)^{-1}$, one obtains $S_\text{eff}\{\pi,A\}=S_\text{eff}\{0,\gtr{U(\pi)^{-1}}A\}$, which ensures that the fields $\pi^a,A^i_\mu$ only appear in a specific combination,
\begin{align}
\label{Bphidef}
\gtr{U(\pi)^{-1}}{A_\mu}&=U(\pi)^{-1}(A_\mu+\imag\partial_\mu)U(\pi)\\
\notag
&\equiv\phi^a_\mu(\pi) T_a+B^\alpha_\mu(\pi) T_\alpha=\phi_\mu(\pi)+B_\mu(\pi).
\end{align}
The broken and unbroken components transform in turn as
\begin{equation}
\gtr\g{\phi_\mu}=\h\phi_\mu\h^{-1},\qquad
\gtr\g{B_\mu}=\h B_\mu\h^{-1}+\imag\h\partial_\mu\h^{-1},
\label{Bphitransfo}
\end{equation}
where $\h$ is given by Eq.~\eqref{pitransfo}. The effective Lagrangian can be split into two parts, $\La_\text{eff}[\phi,B]=\La_\text{inv}[\phi,B]+\La_\text{CS}[B]$~\cite{Leutwyler:1993iq}. The part $\La_\text{inv}$ is strictly invariant under the unbroken gauge transformation~\eqref{Bphitransfo} and can therefore be constructed out of covariant constituents: $\phi_\mu$, $G_{\mu\nu}\equiv\partial_\mu B_\nu-\partial_\nu B_\mu-\imag[B_\mu,B_\nu]$, and their covariant derivatives; see Ref.~\cite{EFTpaper} for more details. The part $\La_\text{CS}$ depends solely on the gauge field $B^\alpha_\mu$ and is quasi-invariant; this is the advertised CS Lagrangian.

The spectrum of NG bosons as well as their dominant interactions at low energy are determined by the leading-order Lagrangian with up to two derivatives, which we find to be
\begin{align}
\label{LOlag}
\La^\text{LO}_\text{eff}&=e^\mu_\alpha B^\alpha_\mu+e^\mu_a\phi^a_\mu+\tfrac12g^{\mu\nu}_{ab}\phi^a_\mu\phi^b_\nu\\
\notag
&=-e^\mu_i\omega^i_a\partial_\mu\pi^a+e^\mu_j\nu^j_iA^i_\mu+\tfrac12g^{\mu\nu}_{ab}\omega^a_c\omega^b_dD_\mu\pi^c D_\nu\pi^d.
\end{align}
The couplings $e^\mu_i$ and $g^{\mu\nu}_{ab}$ are invariant tensors of $H$ (and likewise of the spacetime symmetry), that is, $e^\mu_i f^i_{\alpha j}=0$ and $g^{\mu\nu}_{cb}f^c_{\alpha a}+g^{\mu\nu}_{ac}f^c_{\alpha b}=0$; $f^i_{jk}$ are the structure constants of $G$. The functions $\omega^i_a(\pi)$ and $\nu^j_i(\pi)$ in Eq.~\eqref{LOlag} are given by $U(\pi)$,
\begin{equation}
\omega^i_aT_i\equiv-\imag U^{-1}(\partial U/\partial\pi^a),\qquad
\nu^j_iT_j\equiv U^{-1}T_iU.
\label{MCdef}
\end{equation}
Finally, $D_\mu\pi^a\equiv\partial_\mu\pi^a-A^i_\mu h^a_i(\pi)$ is the covariant derivative of the NG field, where $h^a_i(\pi)$ defines an infinitesimal shift of the NG field under the transformation $\g=e^{\imag\epsilon^iT_i}$ in Eq.~\eqref{pitransfo}.

Assuming rotational invariance, $e^\mu_i=e_i\delta^{\mu0}$~\cite{Watanabe:2013uya,Watanabe:2014fva}. Moreover, $g^{\mu\nu}_{ab}\phi^a_\mu\phi^b_\nu=\bar g_{ab}\phi^a_0\phi^b_0-g_{ab}\phi^a_r\phi^b_r$ where $r$ is a spatial vector index \footnote{In two dimensions, another term is allowed: $\bar{\bar g}_{ab}\epsilon^{rs}\phi^a_r\phi^b_s$~\cite{Watanabe:2014fva}.}. With the particular choice $U(\pi)=e^{\imag\pi^aT_a}$, one then finds by a power expansion in $\pi^a$ that
\begin{equation}
\La_\text{eff}^\text{LO}=\tfrac12e_i f^i_{ab}\partial_0\pi^a\pi^b+e_iA^i_0+\tfrac12g^{\mu\nu}_{ab}D_\mu\pi^aD_\nu\pi^b+\dotsb.
\label{LOlagexp}
\end{equation}
Every pair $T_a,T_b$ such that $\frac1{V}\langle0|[\hat T_a,\hat T_b]|0\rangle=\imag f^i_{ab}e_i\neq0$ ($V$ being spatial volume) gives rise to a canonically conjugate pair of variables, hence \emph{one} type-B NG boson~\cite{Watanabe:2012hr} with, as a rule, quadratic dispersion relation. The remaining $\pi^a$s excite one type-A NG boson each, with a typically linear dispersion.


\section{Chern--Simons terms}
\label{sec:CS}

Eq.~\eqref{LOlag} features the simplest example of a CS term: $e^\mu_\alpha B^\alpha_\mu$. We will now show how to construct such terms systematically. The gauge current, defined by $J^\mu_\alpha[B]\equiv\delta S_\text{CS}\{B\}/\delta B^\alpha_\mu$, satisfies the current conservation, $\partial_\mu J^\mu_\alpha+f^\gamma_{\alpha\beta}J^\mu_\gamma B^\beta_\mu=0$, and transforms under $\h\in H$ with infinitesimal parameters $\epsilon^\alpha$ as $\delta J^\mu_\alpha=-f^\gamma_{\alpha\beta}J^\mu_\gamma\epsilon^\beta$. Due to the latter, the current can be built solely out of covariant constituents: $G^\alpha_{\mu\nu}$ and its covariant derivatives. The Lagrangian is in turn reconstructed using
\begin{equation}
\La_\text{CS}[B]=\int_0^1\dd t\,B^\alpha_\mu J^\mu_\alpha[tB].
\label{master}
\end{equation}
It is easy to solve the covariance and conservation constraints on $J^\mu_\alpha$ at the lowest orders in the gradient expansion. Up to order three, the only solutions are a constant, $e^\mu_\alpha$, and $c^{\mu\nu\lambda}_{\alpha\beta}G^\beta_{\nu\lambda}$. Integration indicated in Eq.~\eqref{master} then leads to
\begin{equation}
\begin{split}
\La_\text{CS}^{(1)}={}&e^\mu_\alpha B^\alpha_\mu,\qquad\text{where}\quad e^\mu_\gamma f^\gamma_{\alpha\beta}=0,\\
\La_\text{CS}^{(3)}={}&c^{\mu\nu\lambda}_{\alpha\beta}B^\alpha_\mu(\partial_\nu B^\beta_\lambda+\tfrac13f^\beta_{\gamma\delta}B^\gamma_\nu B^\delta_\lambda),\\
&\text{where}\quad c^{\mu\nu\lambda}_{\gamma\beta}f^\gamma_{\delta\alpha}+c^{\mu\nu\lambda}_{\alpha\gamma}f^\gamma_{\delta\beta}=0;
\end{split}
\label{CSterms}
\end{equation}
$c^{\mu\nu\lambda}_{\alpha\beta}$ is antisymmetric in $\mu,\nu,\lambda$ and symmetric in $\alpha,\beta$. These are all CS terms up to order four in derivatives~\footnote{For a proof of absence of CS terms at order four, see Ref.~\cite{EFTpaper}}. Lorentz invariance only allows $\La_\text{CS}^{(3)}$ in three spacetime dimensions, where $c^{\mu\nu\lambda}_{\alpha\beta}=\epsilon^{\mu\nu\lambda}c_{\alpha\beta}$~\cite{Leutwyler:1993iq}. Without Lorentz invariance, $\La_\text{CS}^{(1)}$ is allowed, too, as well as another option in four spacetime dimensions, $c^{\mu\nu\lambda}_{\alpha\beta}=\epsilon^{\kappa\mu\nu\lambda}c_{\kappa,\alpha\beta}$. From now on we will assume that only $e^0_\alpha\equiv e_\alpha$ and $c_{0,\alpha\beta}\equiv c_{\alpha\beta}$ are nonzero.

The expression~\eqref{CSterms} for the CS terms is valid for arbitrary, albeit local, parametrization $\pi^a$ of $G/H$ around its origin. This is sufficient for the physics of NG bosons, yet a globally valid parametrization may be needed even at low energy. For instance, even a weak field $A^i_\mu$ may sweep the ground state through the whole coset space, giving rise to a Berry phase, corresponding to $\La^{(1)}_\text{CS}$~\cite{Watanabe:2014fva,Aitchison:1986qn}. A globally valid matrix expression for the CS terms can be achieved for symmetric coset spaces, that is, such $G$ and $H$ that admit an automorphism $\mathcal{R}$ under which $\mathcal{R}(T_\alpha)=T_\alpha$ and $\mathcal{R}(T_a)=-T_a$, and thus $f^a_{bc}=0$. Setting $U(\pi)=e^{\imag\pi^aT_a}$, there is a field variable that transforms linearly under the whole group $G$~\cite{Callan:1969sn},
\begin{equation}
\Sigma(\pi)\equiv U(\pi)^2,\qquad
\Sigma(\pi'(\pi,\g))=\g\Sigma(\pi)\mathcal{R}(\g)^{-1}.
\end{equation}
Next, use the fact that for semisimple Lie algebras the Killing form is nondegenerate to define the dual vector $e^\alpha$ by $e_\alpha=e^\beta\tr(T_\alpha T_\beta)$. The densities $e_\alpha$ can then be encoded in the matrix variable $Q(\pi)\equiv U(\pi)(e^\alpha T_\alpha) U(\pi)^{-1}=e^\alpha\nu^i_\alpha(\pi)T_i$. Since $e^\alpha T_\alpha$ commutes with all generators of $H$, this likewise transforms linearly under the whole $G$: $Q(\pi')=\g Q(\pi)\g^{-1}$.

In order to express $\La^{(1)}_\text{CS}$ in terms of these linearly transforming variables, we have to extend the domain on which the fields $\pi^a$ are defined~\cite{Witten:1983tw}. With a suitable boundary condition on $\pi^a$, the time manifold can be compactified to a circle, $S^1$. Provided that $G/H$ is simply connected, there is an interpolation $\tilde\pi^a(\tau,x)$ for $\tau\in[0,1]$ such that $\tilde\pi^a(0,x)=0$ and $\tilde\pi^a(1,x)=\pi^a(x)$. The coordinates $\tau,t$ then define a unit disk, $D^2$, and the action associated with $\La^{(1)}_\text{CS}$ becomes
\begin{equation}
\begin{split}
S^{(1)}_\text{CS}={}&\frac\imag4\int\dd^d\bm x\int_{D^2}\epsilon^{mn}\tr(Q\partial_m\Sigma\partial_n\Sigma^{-1})\\
&+\int\dd t\,\dd^d\bm x\,\tr(QA_0).
\end{split}
\label{CS1symmetric}
\end{equation}
Here $m,n$ label coordinates on $D^2$ ordered so that $\epsilon^{\tau t}=1$. This matrix form of $S^{(1)}_\text{CS}$, suitable for practical applications, generalizes expressions found before for various specific systems such as ferromagnets~\cite{Volovik:1987li}, $\gr{SU}(N)$ ferromagnets~\cite{Affleck:1988wz,*Wiegmann:1988qn,*Read:1989jy}, superfluid Helium~\cite{Goff:1988qq}, or $\gr{SO(5)}$ spin chains~\cite{Lee:2010ni}.

Similar reasoning applies to the order-three CS term. We use the factorization $c^{\mu\nu\lambda}_{\alpha\beta}=\epsilon^{\mu\nu\lambda}c_{\alpha\beta}$ valid in three spacetime dimensions~\footnote{The argument below applies without changes to four spacetime dimensions, as long as the base manifold $S^3$ is interpreted as space and an additional integration over time is included.} and represent the invariant coupling $c_{\alpha\beta}$ by a matrix $\Xi_0$ so that $c_{\alpha\beta}=\tr(\Xi_0T_\alpha T_\beta)$. Such $\Xi_0$ certainly exists when $H$ is semisimple; see also the discussion of a concrete example in Sec.~\ref{subsec:QHF}. The variable $\Xi(\pi)\equiv U(\pi)\Xi_0 U(\pi)^{-1}$ now transforms linearly just like $Q(\pi)$ and allows us to rewrite the part of $S^{(3)}_\text{CS}$, independent of the external gauge field, in the simple matrix form
\begin{equation}
S^{(3)}_\text{CS}\Bigr|_{A=0}=-\frac1{16}\int_{D^4}\epsilon^{k\ell mn}\tr(\Xi\partial_k\Sigma\partial_\ell\Sigma^{-1}\partial_m\Sigma\partial_n\Sigma^{-1}).
\label{CS3symmetric}
\end{equation}
Here we have assumed that the spacetime can be compactified to $S^3$ and that $\pi_3(G/H)=0$ so that the NG fields can be smoothly extended to $\tilde\pi^a(\tau,x)$, defined on the four-disk, $D^4$. The coordinates on $D^4$ are ordered so that $\epsilon^{\tau123}=1$.

A derivation of Eqs.~\eqref{CS1symmetric} and \eqref{CS3symmetric} together with their generalization to arbitrary, not necessarily symmetric, coset spaces is provided in Ref.~\cite{supplement}.


\subsection{Topological nature of Chern--Simons terms}
\label{subsec:topology}

The CS terms are singled out by our construction, but what makes them special physics-wise? First, \emph{some} of the CS couplings may be quantized, depending on the topology of spacetime and of the coset space $G/H$~\cite{Witten:1983tw,supplement}. Due to the extra spatial integral in Eq.~\eqref{CS1symmetric}, $e_\alpha$ can only be quantized in a finite space volume $V$. Likewise, $c_{\alpha\beta}$ is quantized in three spacetime dimensions, or possibly in four dimensions provided the time volume is finite. In any case, the topological nature of the CS terms is expected to manifest in the non-renormalization of their couplings under quantum corrections~\cite{Dunne:1998qy}.

The order-one CS term has another notable consequence: its contribution to canonical momentum density, $P_r=e_\alpha B^\alpha_r$, is not invariant under the internal symmetry group $G$. This is known in ferromagnets as the linear momentum problem~\cite{Haldane:1986zz}, which is also related to the topology of the coset space~\cite{Papanicolaou:1990sg,*Floratos:1992hf,*Nair:2003vm}. In some systems such as ferromagnetic metals~\cite{Volovik:1987li} or superfluid Helium~\cite{Goff:1988qq}, the resolution of this paradox is through the presence of gapless fermionic degrees of freedom, which makes the EFT for the NG modes alone incomplete, or even ill-defined by inducing nonlocal terms in the action~\cite{Nayak:current}. Our EFT framework makes it clear that the phenomenon is general, suggesting that type-B NG modes associated with unbroken charge in the ground state are always accompanied by other (whether NG or non-NG) gapless modes.

Another outstanding feature of \emph{all} CS terms is their insensitivity to local deformations of the system. Consider a medium whose microscopic properties vary in space. Such a variation can be taken into account in $\La_\text{inv}$ without violating $G$-invariance by making the couplings coordinate-dependent. This is in general not possible for the quasi-invariant terms though, as arbitrary coordinate dependence of, say, $e_\alpha$ would spoil the $G$-invariance of $S^{(1)}_\text{CS}$, and likewise for the other CS terms. The most general form of the order-one CS term compatible with the internal symmetry is $e^\mu_\alpha B^\alpha_\mu$, where $e^\mu_\alpha$ is now a function of coordinates that is invariant under $H$ and satisfies the conservation condition $\partial_\mu e^\mu_\alpha=0$.

Finally, the CS terms cannot be written in a way that preserves both manifest locality and gauge invariance. Eq.~\eqref{CSterms} obviously sacrifices the latter. This can be fixed by interpolating the fields $A^i_\mu$ to the extended base manifold, $D^2$ or $D^4$, along with $\pi^a$. The resulting expression, however, obscures locality, being a sum of terms each of which depends on the interpolation $\tilde\pi^a$ rather than on the physical values of $\pi^a$~\cite{supplement}.


\subsection{Discrete symmetries}
\label{subsec:symmetries}

Both the $e_a$ and the (CS) $e_\alpha$ term in Eq.~\eqref{LOlag} break explicitly certain discrete symmetry (not to be confused with time reversal~\cite{Watanabe:2014fva,Roman:1999ro}). To that end, note that the generators can be chosen so that all those with a nonzero vacuum expectation value are diagonal~\cite{Watanabe:2011ec}. Now set $U(\pi)=e^{\imag\pi^aT_a}$ and define a ``charge conjugation'' $\mathcal{C}$ by
\begin{equation}
\mathcal{C} U(\pi)\equiv U(\pi)^*=U(\pi)^{-1T}.
\label{Cparity}
\end{equation}
One easily finds that $\mathcal{C}\omega=-\omega^T$; gauge covariance is thus preserved by defining $\mathcal{C}A_\mu=-A^T_\mu$. As a rule, the two-derivative Lagrangian in Eq.~\eqref{LOlag} preserves $\mathcal{C}$; when the NG fields are irreducible under $H$, this follows from $g_{ab}\phi^a_\mu\phi^b_\nu\propto\tr(\phi_\mu\phi_\nu)$. On the other hand, $e_\alpha B^\alpha_0=\tr(e^\alpha T_\alpha B_0)$ changes sign under $\mathcal{C}$ since $e^\alpha T_\alpha$ is by assumption diagonal. The same argument applies to the invariant term $e_a\phi^a_0$.

$\mathcal C$ is an accidental symmetry of the two-derivative terms, similar to the intrinsic parity in the chiral perturbation theory, defined as $\pi^a\to-\pi^a$~\cite{Beane:1998xa,*Scherer:2002tk}. Its breaking may lead to certain ``anomalous'' processes such as magnon decay into photons in two-dimensional (anti)ferromagnets~\cite{Bar:2004bw}. The intrinsic parity itself is preserved, at least for symmetric coset spaces, by the CS terms since it leaves invariant their building block, $B^\alpha_\mu$.


\subsection{Chern--Simons interactions of Nambu--Goldstone bosons}
\label{subsec:interactions}

The physical importance of the order-one terms in the Lagrangian~\eqref{LOlag} is clear: they determine the dispersion relations of NG bosons as well as their leading interactions. On the contrary, the implications of $\La^{(3)}_\text{CS}$ for the NG bosons are subtle. \emph{Suppose} first that there is a $G$-invariant tensor coupling $C_{ij}$ such that $C_{\alpha\beta}=c_{\alpha\beta}$. Any $G$-invariant $C_{ij}$ satisfies the identity (using the notation $\omega^i_\mu\equiv\omega^i_a\partial_\mu\pi^a$)
\begin{align}
\label{auxCStransfo}
\epsilon^{\mu\nu\lambda}&C_{\alpha\beta}B^\alpha_\mu(\partial_\nu B^\beta_\lambda+\tfrac13f^\beta_{\gamma\delta}B^\gamma_\nu B^\delta_\lambda)\\
\notag
={}&\epsilon^{\mu\nu\lambda}C_{ij}A^i_\mu(\partial_\nu A^j_\lambda+\tfrac13f^j_{k\ell}A^k_\nu A^\ell_\lambda)+\tfrac16\epsilon^{\mu\nu\lambda}C_{ij}f^j_{k\ell}\omega^i_\mu\omega^k_\nu\omega^\ell_\lambda\\
\notag
&-\epsilon^{\mu\nu\lambda}\bigl(C_{a\alpha}\phi^a_\mu G^\alpha_{\nu\lambda}+C_{ab}\phi^a_\mu D_\nu\phi^b_\lambda+\tfrac13C_{ai}f^i_{bc}\phi^a_\mu\phi^b_\nu\phi^c_\lambda\bigr)
\end{align} 
up to a surface term. This allows us to rewrite $\La^{(3)}_\text{CS}$ as a sum of: (i) a CS term for $A^i_\mu$ alone plus a $\theta$-term [second line of Eq.~\eqref{auxCStransfo}]; (ii) invariant terms from $\La_\text{inv}$ (last line). Therefore, $\La^{(3)}_\text{CS}$ does not induce any interactions among the NG bosons. In ferromagnets $\La^{(3)}_\text{CS}$ is known as the Hopf term~\cite{Wilczek:1983cy,*Volovik:2003fe}. 

When $H$ is simple, $c_{\alpha\beta}$ is proportional to $\tr(T_\alpha T_\beta)$ by Schur's lemma~\cite{Georgi:1982jb}; we can then define $C_{ij}$ by $\tr(T_iT_j)$. A necessary condition for $\La^{(3)}_\text{CS}$ to trigger interactions among NG bosons is therefore that $H$ is not simple. Expanding in powers of $\pi^a$ then yields
\begin{equation}
\La_\text{CS}^{(3)}\Bigr|_{A=0}=\tfrac14\epsilon^{\mu\nu\lambda}c_{\alpha\beta}f^\alpha_{ab}f^\beta_{cd}\pi^a\partial_\mu\pi^b\partial_\nu\pi^c\partial_\lambda\pi^d+\dotsb.
\label{WZW}
\end{equation}
While formally reminiscent of the WZ term in the chiral perturbation theory, this interaction, hitherto unnoticed, does \emph{not} arise from anomalous microscopic dynamics. For an example, consider the class of symmetry-breaking patterns $G_1\times G_2\to H_1\times H_2$, where $H_i\subset G_i$.  The fields $\phi^a_\mu,B^\alpha_\mu$ then split into separate contributions from each $G_i/H_i$. Provided there is no singlet of $H$ among the broken generators, the two sets of NG fields enter separately both the leading-order Lagrangian~\eqref{LOlag} and the order-three invariant one~\footnote{At order three, the available operators for $\La_\text{inv}$ are $\phi^a_\mu\phi^b_\nu\phi^c_\lambda$, $\phi^a_\mu D_\nu\phi^b_\lambda$ and $\phi^a_\mu G^\alpha_{\nu\lambda}$.}. If, in addition, both $H_i$ contain a $\gr{U(1)}$ factor, a coupling $c_{\alpha\beta}$ mixing the two is compatible with $H$-invariance. Eq.~\eqref{WZW} then provides the \emph{leading} interaction among NG bosons from the two coset spaces $G_i/H_i$. A symmetry-breaking pattern of the above type occurs for instance in the A-phase of liquid Helium~\cite{Vollhardt:1990vw}. However, the broken symmetry in this case includes spatial rotations, not covered by the present paper, which is concerned exclusively with internal symmetries.
 

\section{Examples}
\label{sec:example}

\subsection{Ferromagnets}
\label{subsec:ferromagnets}

Let us illustrate the general arguments on examples, starting with the simplest case of a spin-$\frac12$ ferromagnet. As pointed out in Ref.~\cite{Frohlich:1993gs}, the nonrelativistic Pauli equation in presence of an electromagnetic field features an $G=\gr{SU(2)}_s\times\gr{U(1)}_{em}$ \emph{gauge} invariance. Here the $\gr{SU(2)}_s$ factor represents electromagnetic interactions of spin and the associated gauge potentials $\vec A_\mu$ are given by the electric and magnetic field intensities. The $\gr{U(1)}_{em}$ factor, on the other hand, describes coupling of electric charge to the electromagnetic gauge potential $A^{em}_\mu$. Spontaneous magnetization in the ground state of a ferromagnet (chosen without loss of generality to point in the $z$ direction) breaks the symmetry to $H=\gr{U(1)}_s\times\gr{U(1)}_{em}$.

It is common to describe the magnetization by a unit vector $\vec n$, related to our general notation by $\vec\sigma\cdot\vec n=\Sigma\sigma_3=U\sigma_3U^{-1}$, where $\vec\sigma$ is the vector of Pauli matrices. The order-one CS term~\eqref{CS1symmetric} then takes the usual form~\cite{Volovik:1987li}
\begin{equation}
S^{(1)}_\text{CS}=M_0\int\dd^d\bm x\int_{D^2}\vec n\cdot(\partial_t\vec n\times\partial_\tau\vec n)+M_0\int\dd t\,\dd^d\bm x\,\vec n\cdot\vec A_0,
\end{equation}
where $M_0$ is the spin density in the ground state. The first term is responsible for the Larmor precession of spin as described by the Landau--Lifschitz equation~\cite{Leutwyler:1993gf}. The second term gives the Zeeman coupling of the magnetization to the magnetic field $\vec B=\vec A_0/\mu$, $\mu$ being the magnetic moment.

Let us inspect possible order-three CS terms, restricting from now on to $d=2$. Since the unbroken subgroup $H$ has two $\gr{U(1)}$ factors, there are three different terms, corresponding to the independent entries of the (symmetric) matrix $c_{\alpha\beta}$. First, the $\gr{U(1)}_s$ term $B^3\wedge\dd B^3$ can be by Eq.~\eqref{auxCStransfo} absorbed into a CS term for $\vec A_\mu$ alone. It does not affect the perturbative dynamics of NG bosons, as is clear from Eq.~\eqref{WZW}. It is relevant for topologically nontrivial spin configurations though: the $\theta$-term on the second line of Eq.~\eqref{auxCStransfo} is the Hopf term. Second, the $\gr{U(1)}_{em}$ term $A^{em}\wedge\dd A^{em}$ is independent of the NG fields, as the Abelian gauge field $A^{em}_\mu$ is unaffected by the field redefinition~\eqref{Bphidef}; it describes the Hall effect.

The most interesting is the mixed CS term $A^{em}\wedge\dd B^3$. By Eq.~\eqref{CSterms}, this connects $A^{em}_\mu$ to magnons through the current $\epsilon^{\mu\nu\lambda}\partial_\nu B^3_\lambda=\frac12\epsilon^{\mu\nu\lambda}G^3_{\nu\lambda}$~\cite{supplement,Ray:elmag}. This is a topological current whose integral charge is, for vanishing $\vec A_\mu$, proportional to the topological winding number, $\frac1{8\pi}\int\dd^2\bm x\,\epsilon^{rs}\vec n\cdot(\partial_r\vec n\times\partial_s\vec n)$. The associated effective coupling can therefore be interpreted as the electric charge of a topological soliton, called the baby-skyrmion. In ferromagnets, the $\mathcal C$ conjugation~\eqref{Cparity} acts as a reflection in the tangent plane to $G/H\simeq S^2$ at $\pi=0$, and is equivalent to the inversion $\vec n\to-\vec n$ up to a finite $\gr{SU(2)}_s$ rotation. Hence both the order-one CS term and the mixed order-three CS term are $\mathcal{C}$-odd. The latter gives the leading contribution to the magnon decay into a pair of photons~\cite{Bar:2004bw}.


\subsection{Quantum Hall ferromagnets}
\label{subsec:QHF}

An intriguing generalization of the above simple example is provided by quantum Hall ferromagnets, whether realized by multilayered ferromagnets~\cite{EzawaQHF,*HamaQHF} or by Landau-level degeneracy in graphene~\cite{Yang:2006prb}. Assuming first for simplicity exact degeneracy we have $G=\gr{SU}(N)$, where $N$ is the total number of levels. The ferromagnetic order parameter can be viewed as a Hermitian matrix $\Phi$ transforming as $\Phi\to\g\Phi\g^{-1}$ under $G$. In the ground state, $\Phi$ reduces to
\begin{equation}
\langle\Phi\rangle=\text{diag}(\underbrace{\lambda_1,\dotsc,\lambda_1}_{M\times},\underbrace{\lambda_2,\dotsc,\lambda_2}_{(N-M)\times}),
\end{equation}
breaking the symmetry down to $H=\gr{S[U}(M)\times\gr{U}(N-M)]$, where $M$ is the filling factor, supposed here to be an integer. The coset space $G/H$ is symmetric, the automorphism $\mathcal{R}$ being given by a matrix $R\equiv\text{diag}(+1,\dotsc,+1,-1,\dotsc,-1)$. This allows us to define a unitary Hermitian matrix variable $\mathcal{N}\equiv\Sigma R=URU^{-1}$~\footnote{In Ref.~\cite{Bar:2004bw}, the special case $M=1$ was investigated using the projector variable $P=\frac12(\openone+\mathcal{N})$.}; this generalizes the matrix variable $\vec\sigma\cdot\vec n$, used above for spin-$\frac12$ ferromagnets, which correspond to $N=2$ and $M=1$.

The coset space $G/H$---the Grassmannian~\cite{Nakahara}---has dimension $2M(N-M)$, hence the ferromagnetic ground state supports $M(N-M)$ type-B magnon excitations. Their dynamics is driven by the order-one CS term. This is specified by a single effective coupling, corresponding to the sole $\gr{U(1)}$ generator of $H$, proportional to $-\frac N2R+(M-\frac N2)\openone$. According to Eq.~\eqref{CS1symmetric}, the order-one CS term therefore reads
\begin{equation}
S^{(1)}_\text{CS}\Bigr|_{A=0}=\frac{\imag M_0}4\int\dd^2\bm x\int_{D^2}\epsilon^{mn}\tr(\mathcal N\partial_m\mathcal N\partial_n\mathcal N),
\end{equation}
where the parameter $M_0$ again stands for the size of magnetization in the ground state.

For $2\leq M\leq N-2$, the coupling $c_{\alpha\beta}$ encodes three parameters, one of which can be eliminated via Eq.~\eqref{auxCStransfo}. The remaining two parametrize the matrix $\Xi_0$, whose most general form compatible with the unbroken symmetry is $\Xi_0=cR+d\openone$. It is now straightforward, albeit a bit tedious, to evaluate the CS term~\eqref{CS3symmetric} in terms of $\mathcal N$,
\begin{equation}
S^{(3)}_\text{CS}\Bigr|_{A=0}=-\frac{c}{16}\int_{D^4}\epsilon^{k\ell mn}\tr(\mathcal{N}\partial_k\mathcal{N}\partial_\ell\mathcal{N}\partial_m\mathcal{N}\partial_n \mathcal{N}).
\label{QHf}
\end{equation}
This form was derived in the special case $M=1$ in Ref.~\cite{Bar:2004bw}. Since the matrix $R$ is real and diagonal, the $\mathcal C$ conjugation~\eqref{Cparity} amounts to $\mathcal N\to\mathcal N^T$. It immediately follows that $S^{(1)}_\text{CS}$ and $S^{(3)}_\text{CS}$ is $\mathcal C$-odd and $\mathcal C$-even, respectively.

Consider now a quantum Hall ferromagnet in graphene, where approximate spin and valley symmetries combine into $G=\gr{SU(4)}$. At zero doping, the lowest Landau level is half-filled, that is, $M=2$. The $\gr{SU(2)}_{s,v}$ factors of $H$ can be identified with spin and valley (pseudospin) rotations. The interactions of the associated NG bosons are described by Eq.~\eqref{QHf}. In reality, the $\gr{SU(4)}$ symmetry is only approximate. In the quantum Hall regime of graphene, the most dominant explicit symmetry breaking effects are the Zeeman splitting and the Kekul\'e-type lattice distortion \cite{RevModPhys.83.1193}. While the former breaks $\gr{SU(2)}_s$ and spin-polarizes the system, the latter breaks $\gr{SU(2)}_v$. Provided that the Zeeman splitting is negligible~\footnote{Unfortunately, this is difficult to justify in current experiments.}, the symmetry-breaking pattern reduces to $\gr{SU(2)}_s\to\gr{U(1)}_s$, which is just the well-known case of a spin ferromagnet. Thus the coupling of the graphene quantum Hall ferromagnet  to electromagnetism in this particular regime is identical with the case of a spin ferromagnet already discussed above. 


\section{Conclusions}
\label{sec:conclusions}

In this paper, we have provided a general classification of quasi-invariant Lagrangians for NG bosons in many-body systems, without assuming specific spacetime symmetry. In addition to the practically useful explicit expressions~\eqref{CSterms}, \eqref{CS1symmetric} and \eqref{CS3symmetric} for the ensuing CS terms, we would like to stress the simplicity of the approach advocated here, as compared to existing literature~\cite{DHoker:1995it}. Using the ideas of general coordinate invariance, relativistic or not~\cite{Son:2005rv,*Son:2008ye}, we expect it to readily generalize to broken spacetime symmetries. In combination with the formalism proposed recently in Ref.~\cite{Brauner:2014jaa}, the method could therefore offer a novel EFT approach to systems such as solids~\cite{Leutwyler:1996er}, supersolids~\cite{Son:2005ak}, or exotic superfluids~\cite{Hoyos:2013eha}. This generalization would also allow one to discuss mixing of sound with other NG modes. We plan to address these points in our future work.


\section*{Acknowledgments}

One of us (T.B.) gratefully appreciates collaboration with Jens O.~Andersen, Christoph P.~Hofmann and Aleksi Vuorinen on a closely related project~\cite{EFTpaper}. Our understanding was shaped by discussions and correspondence with Paulo Bedaque, Heinrich Leutwyler, Jouko Mickelsson, Riccardo Rattazzi, Mikko Sainio, Grigori Volovik, Haruki Watanabe and Uwe-Jens Wiese. The work of T.B.~was supported by the Academy of Finland, Grant No.~273545, and by the Austrian Science Fund (FWF), Grant No.~M~1603-N27. He also acknowledges the hospitality of IFT UAM-CSIC, afforded through the Centro de Excelencia Severo Ochoa Program, Grant No.~SEV-2012-0249. The work of S.M.~was supported by the US DOE Grant No.~DE-FG02-97ER-41014.


\bibliography{references}


\clearpage

\section*{Supplemental material}

In this Supplemental Material, we provide the reader with some details which either are rather technical and thus probably of interest only to the specialists, or are not essential for understanding the key steps of our construction, merely offering additional insight. In particular, we: (i) describe a differential-geometric construction of the Chern--Simons (CS) terms which generalizes Eqs.~(\ref*{CS1symmetric}) and (\ref*{CS3symmetric}) of the main text; (ii) use this construction to derive quantization conditions for couplings in the CS terms; (iii) discuss to some extent the associated topological currents.


\subsection*{Wess--Zumino--Witten construction}

The expression~(\ref*{CSterms}) in the main text for the CS terms is completely general, yet not particularly transparent in that it does not make invariance under the group $G$ manifest. Fortunately, an alternative exists which remedies this unwanted feature. To that end, it is suitable to use exterior calculus. The composite gauge field $\gtr{U(\pi)^{-1}}A_\mu=\phi_\mu(\pi)+B_\mu(\pi)$ thereby becomes a 1-form, $\gtr{U(\pi)^{-1}}A=-\omega(\pi)+\mathcal{A}(\pi)$, where
\begin{equation}
\omega\equiv-\imag U^{-1}\dd U=T_i\omega^i_a\dd\pi^a
\end{equation}
is the Maurer--Cartan (MC) form and we introduced the shorthand notation $\mathcal{A}\equiv T_j\nu^j_iA^i_\mu\dd x^\mu$. Note that, strictly speaking, what appears in $\gtr{U(\pi)^{-1}}A$ is the pull-back of the MC form by the Nambu--Goldstone (NG) field, $\pi^*\omega=T_i\omega^i_a\partial_\mu\pi^a\dd x^\mu$. We will take the liberty to identify this with $\omega$ itself as there is no danger of confusion.

Using the MC structure equation, $\dd\omega^i=\tfrac12f^i_{jk}\omega^j\wedge\omega^k$, and the invariance condition $e_\gamma f^\gamma_{\alpha\beta}=0$, one immediately observes that $\dd(e_\alpha\omega^\alpha)=\frac12e_\alpha f^\alpha_{ab}\omega^a\wedge\omega^b$. This 2-form is invariant under global $G$ transformations, unlike $e_\alpha\omega^\alpha$, and hence $\La^{(1)}_\text{CS}$, itself. It can be used to rewrite the action in a form manifestly globally invariant by promoting the NG fields $\pi^a(x)$ to the fields $\tilde \pi^a(\tau,x)$, defined on the extended base manifold $D^2$. (This step assumes that the coset space $G/H$ is simply connected.) The Stokes theorem then implies that
\begin{equation}
S^{(1)}_\text{CS}=-\frac12e_\alpha f^\alpha_{ab}\int\dd^d\bm x\int_{D^2}\omega^a\wedge\omega^b+e_\alpha\int\dd t\,\dd^d\bm x\,\mathcal{A}^\alpha_0.
\label{WZ1}
\end{equation}
In the special case of a symmetric coset space, one can take advantage of the existence of the linearly transforming variable $\Sigma$. From its definition, one finds by a short manipulation
\begin{equation}
\omega^aT_a=-\tfrac\imag2U^{-1}(\dd\Sigma)U^{-1}=+\tfrac\imag2U(\dd\Sigma^{-1})U.
\end{equation}
Upon converting the 2-form $e_\alpha f^\alpha_{ab}\omega^a\wedge\omega^b$ into a trace of a product of matrices, this recovers Eq.~(\ref*{CS1symmetric}) of the main text.

The order-three CS term can be dealt with in the same manner. We again use the factorization $c^{\mu\nu\lambda}_{\alpha\beta}=\epsilon^{\mu\nu\lambda}c_{\alpha\beta}$, valid in three spacetime dimensions, and rewrite the Lagrangian $\La^{(3)}_\text{CS}$ as a 3-form, $c_{\alpha\beta}B^\alpha\wedge(\dd B^\beta+\frac13f^\beta_{\gamma\delta}B^\gamma\wedge B^\delta)$. The exterior derivative of this form is $c_{\alpha\beta}G^\alpha\wedge G^\beta$. With the help of the MC equation, the invariance condition $c^{\mu\nu\lambda}_{\gamma\beta}f^\gamma_{\delta\alpha}+c^{\mu\nu\lambda}_{\alpha\gamma}f^\gamma_{\delta\beta}=0$ and the Stokes theorem, one obtains after some manipulation the general expression
\begin{align}
\notag
S^{(3)}_\text{CS}={}&\frac14c_{\alpha\beta}f^\alpha_{ab}f^\beta_{cd}\int_{D^4}\omega^a\wedge\omega^b\wedge\omega^c\wedge\omega^d\\
\notag
&+c_{\alpha\beta}\int_{S^3}\bigl[-f^\alpha_{ab}\omega^a\wedge\omega^b\wedge\mathcal{A}^\beta+f^\alpha_{ab}\omega^a\wedge\mathcal{A}^b\wedge\mathcal{A}^\beta\\
&+\mathcal{A}^\alpha\wedge(\nu^\beta_i\dd A^i+\tfrac13f^\beta_{\gamma\delta}\mathcal{A}^\gamma\wedge\mathcal{A}^\delta)\bigr].
\label{WZ3}
\end{align}
This implicitly assumes that spacetime is compactified to $S^3$ and that $\pi_3(G/H)=0$; the interpolation fields $\tilde\pi^a(\tau,x)$ are thus defined on the disk $D^4$. With the matrix representation for the coupling, $c_{\alpha\beta}=\tr(\Xi_0T_\alpha T_\beta)$, one can rewrite the group prefactor as $c_{\alpha\beta}f^\alpha_{ab}f^\beta_{cd}=-\tr(\Xi_0[T_a,T_b][T_c,T_d])$. This leads to Eq.~(\ref*{CS3symmetric}) in the main text.

Our general expressions maintain manifest \emph{global} invariance under the group $G$. The reader may wonder whether it is possible to keep \emph{gauge} invariance manifest as well. This is indeed the case but, as we now demonstrate, it only comes with the cost of giving up manifest locality. The first step is to interpolate the background fields $A^i_\mu$ to the extended base manifold, $D^2$ or $D^4$ respectively, alongside with $\pi^a$. Everything can now be cast in terms of manifestly covariant objects. Introducing the shorthand notation, $\mathcal{F}\equiv\frac12T_j\nu^j_iF^i_{\mu\nu}\dd x^\mu\wedge\dd x^\nu$, where $F^i_{\mu\nu}\equiv\partial_\mu A^i_\nu-\partial_\nu A^i_\mu+f^i_{jk}A^j_\mu A^k_\nu$ is the field strength of the background fields, and using the MC equation, the curvature 2-form for the composite field $B^\alpha$ acquires the form
\begin{equation}
G^\alpha=-\tfrac12f^\alpha_{ab}\phi^a\wedge\phi^b+\mathcal F^\alpha.
\label{Galpha}
\end{equation}
This immediately leads to the gauge-covariant expression
\begin{equation}
S^{(1)}_\text{CS}=\int\dd^d\bm x\int_{D^2}\left(-\frac12e_\alpha f^\alpha_{ab}\phi^a\wedge\phi^b+e_\alpha\mathcal F^\alpha\right).
\label{naivegauged}
\end{equation}
A similar expression exists for $S^{(3)}_\text{CS}$. As observed above, the whole action including the gauge fields is given by an integral of $c_{\alpha\beta}G^\alpha\wedge G^\beta$ over $D^4$, and we just need to substitute from Eq.~\eqref{Galpha} to obtain
\begin{equation}
\begin{split}
S^{(3)}_\text{CS}={}&c_{\alpha\beta}\int_{D^4}\bigl(\tfrac14f^\alpha_{ab}f^\beta_{cd}\phi^a\wedge\phi^b\wedge\phi^c\wedge\phi^d\\
&-f^\alpha_{ab}\phi^a\wedge\phi^b\wedge\mathcal F^\beta+\mathcal F^\alpha\wedge\mathcal F^\beta\bigr).
\end{split}
\label{naivegaugedCS3}
\end{equation}

Eqs.~\eqref{naivegauged} and \eqref{naivegaugedCS3} look elegant, yet each of the terms therein depends on the interpolation $\tilde\pi^a$ rather than on the physical values of $\pi^a$, defined on $S^1$ or $S^3$. This dependence only cancels in the sum; these expressions therefore sacrifice manifest locality. For instance, in ferromagnets one finds
\begin{equation}
S^{(1)}_\text{CS}=M_0\int\dd^d\bm x\int_{D^2}\vec n\cdot(D_t\vec n\times D_\tau\vec n+\vec F_{\tau t}),
\end{equation}
where $D_\mu\vec n\equiv\partial_\mu\vec n+\vec A_\mu\times\vec n$, which obviously obscures the nature of the interaction of the spin degrees of freedom with an external magnetic field: the ``Zeeman'' coupling is now proportional to $\vec F_{\tau t}$.


\subsection*{Quantization of the couplings}

The extension of the base manifold in Eq.~\eqref{WZ1} leads to an ambiguity in $\tilde\pi^a$. In such a case, the action cannot be uniquely defined. Yet, the physics as given by a functional integral over the dynamical field variables $\pi^a$ can be left unaffected provided that $-\frac12e_\alpha f^\alpha_{ab}\int\dd^d\bm x\int_{S^2}\omega^a\wedge\omega^b$ is quantized in units of $2\pi$. This restricts the physically consistent values of $e_\alpha$ as long as the homotopy group $\pi_2(G/H)$, or more generally the de Rham cohomology group $H^2(G/H)$, is nontrivial. For example, in ferromagnets the combination $M_0V$ is quantized as a consequence of the nontrivial homotopy group $\pi_2(G/H)=\pi_2(S^2)=\mathbb{Z}$. This is equivalent to the quantization of spin, and gives a nontrivial constraint on the possible values of $M_0$ in any finite volume $V$.

Likewise, Eq.~\eqref{WZ3} for $S^{(3)}_\text{CS}$ was constructed assuming $\pi_3(G/H)=0$. The ambiguity in the action is now given by $\frac14c_{\alpha\beta}f^\alpha_{ab}f^\beta_{cd}\int_{S^4}\omega^a\wedge\omega^b\wedge\omega^c\wedge\omega^d$, which should again be quantized in units of $2\pi$. The quantization is governed by the homotopy group $\pi_4(G/H)$.

Which of the couplings in the order-three CS term exactly are quantized is somewhat subtle. In the general case, Eq.~(\ref*{auxCStransfo}) of the main text allows us rewrite a \emph{part} of $\La^{(3)}_\text{CS}$ as a sum of a CS term for $A^i_\mu$ plus a $\theta$-term, and invariant terms from $\La_\text{inv}$. The remainder of $\La^{(3)}_\text{CS}$ enters the Wess--Zumino action~\eqref{WZ3} and its couplings are quantized due to \emph{global} invariance. The couplings absorbed into the CS term for $A^i_\mu$ may be quantized as well; coupling quantization is a well-known feature of the (non-Abelian) Chern--Simons theory. However, invariance of our EFT under large \emph{gauge} transformations is nontrivial. Topologically nontrivial backgrounds may alter the ground state and the low-energy spectrum; saving gauge invariance may then require adding new gapless fermionic degrees of freedom to the EFT, in accord with the index theorem.

For an example, recall the quantum Hall ferromagnet discussed in the main text. For $2\leq M\leq N-2$, the coupling $c$ is quantized due to $\pi_4(G/H)=\mathbb{Z}$. On the other hand, the coupling $d$, entering only interactions of NG bosons with external fields, is not quantized; it corresponds to the coupling of the (Abelian) CS theory for the $\gr{U(1)}$ factor of $H$. In the special case of $M=1$ or $M=N-1$, $c_{\alpha\beta}$ only contains two parameters, one of which can be eliminated in favor of a CS theory for $A^i_\mu$ alone. This can be effectively taken into account by setting $c=d$. In this case, $c$ is not quantized, in agreement with the fact that $\pi_4(G/H)=0$.


\pagebreak
\subsection*{Topological currents}

Our construction of the quasi-invariant Lagrangians is based on first finding all covariant currents and then integrating them in order to obtain the action. It is instructive to get back to this point and inspect the form of the currents; this sheds a different light on the nature of the CS interactions.

The current giving rise to $\La^{(1)}_\text{CS}$ is a trivial constant and we thus focus solely on $\La^{(3)}_\text{CS}$. Using the definitions of the current, $J^\mu_\alpha=\epsilon^{\mu\nu\lambda}c_{\alpha\beta}G^\beta_{\nu\lambda}$, and of the auxiliary field $B^\alpha_\mu$, we find
\begin{equation}
\begin{split}
J^\mu_\alpha&=\epsilon^{\mu\nu\lambda}c_{\alpha\beta}(\partial_\nu B^\beta_\lambda-\partial_\lambda B^\beta_\nu+f^\beta_{\gamma\delta}B^\gamma_\nu B^\delta_\lambda)\\
&=\epsilon^{\mu\nu\lambda}c_{\alpha\beta}(\nu^\beta_iF^i_{\nu\lambda}-f^\beta_{ab}\phi^a_\nu\phi^b_\lambda),
\end{split}
\end{equation}
where we used that the field-strength tensor transforms covariantly. For symmetric coset spaces where the matrix $\Xi_0$ exists, this can be further written in the matrix form
\begin{equation}
\begin{split}
J^\mu_\alpha={}&\epsilon^{\mu\nu\lambda}\tr\bigl[\tfrac\imag2U(\Xi_0T_\alpha)U^{-1}D_\nu\Sigma D_\lambda\Sigma^{-1}\\
&+U(\Xi_0T_\alpha)U^{-1}F_{\nu\lambda}\bigr].
\end{split}
\end{equation}
Using this prescription, one can derive, for every generator $T_\alpha$ of the unbroken subgroup, a current that satisfies the covariant conservation law $\partial_\mu J^\mu_\alpha+f^\gamma_{\alpha\beta}J^\mu_\gamma B^\beta_\mu=0$ without using equations of motion. 

However, generators of $\gr{U(1)}$ subgroups of $H$ are special, since they correspond to currents satisfying an ordinary conservation law, $\partial_\mu J^\mu_\alpha=0$. Such currents are often referred to as the Goldstone--Wilczek (GW) currents, and give rise to a topological quantum number each. Provided that $T_\alpha=\openone$ or at least $\Xi_0T_\alpha=\Xi_0$, such a GW current can be written using previously introduced notation simply as
\begin{equation}
J^\mu_\text{GW}=\epsilon^{\mu\nu\lambda}\tr\bigl(\tfrac\imag2\Xi D_\nu\Sigma D_\lambda\Sigma^{-1}+\Xi F_{\nu\lambda}\bigr).
\end{equation}

Due to their Abelian nature, the GW currents are linear in the gauge field $B^\alpha_\mu$, and therefore the induced CS interaction term becomes merely $\La_\text{CS}=\frac12B^\alpha_\mu J^\mu_{\text{GW},\alpha}$. For example, in ferromagnets the GW current of the unbroken $\gr{U(1)}_s$ symmetry is proportional to $\epsilon^{\mu\nu\lambda}\vec n\cdot(D_\nu\vec n\times D_\lambda\vec n-\vec F_{\nu\lambda})$. The mixed $A^{em}\wedge\dd B^3$ CS term represents the electromagnetic coupling of this current, with the coupling constant being interpreted as the electric charge of the corresponding topologically nontrivial field configuration: the baby-skyrmion.


\end{document}